\documentclass[a4paper,12pt]{article}  
\usepackage{amssymb}
\usepackage{latexsym}
\usepackage[dvips]{graphicx} 
\usepackage{cite}
\newcommand{\be}{\begin{equation}}
\newcommand{\ee}{\end{equation}}
\newcommand{\bea}{\begin{eqnarray}}
\newcommand{\nn}{\nonumber}
\newcommand{\eea}{\end{eqnarray}}

\begin{document}

\begin{titlepage}
\begin{flushright}
hep-th/0010097\\
UA/NPPS-02-2000
\end{flushright}
\begin{centering}
\vspace{.8in}
{\large {\bf Bogoliubov Coefficients of 2D
Charged Black Holes }} \\ 

\vspace{.5in}

{\bf T. Christodoulakis\footnote{tchris@cc.uoa.gr}}, 
{\bf G.A. Diamandis\footnote{gdiamant@cc.uoa.gr}},
{\bf B.C. Georgalas\footnote{vgeorgal@cc.uoa.gr}} {\bf and}
{\bf E.C. Vagenas\footnote{hvagenas@cc.uoa.gr}} \\

\vspace{0.3in}
University of Athens, Physics Department, 
Nuclear and Particle Physics Section, 
Panepistimioupolis, Ilisia GR 157 71, Athens, Greece. 

\end{centering}

\vspace{1in}

\begin{abstract}
We exactly calculate  the  thermal  distribution  and temperature of 
Hawking radiation for a two-dimensional charged  dilatonic black hole after
 it has settled down to an ``equilibrium" state.
  The calculation is carried out using the Bogoliubov coefficients.
  The background of the process is furnished by a preexisting black hole
  and not by collapsing matter as considered by Giddings and Nelson
  for the case of a Schwarzschild black hole. Furthermore,
  the vanishing of the temperature and/or the Hawking radiation
  in the extremal case is obtained as a regular limit of the general case.
\end{abstract}

\end{titlepage}

\newpage

\baselineskip=18pt

\section*{Introduction} 

By 1973, it was well known that black holes had two main properties:
(i) the total surface of the event horizon cannot decrease and (ii) the
surface gravity is constant on the event horizon.
These properties were elements of the classical theory of black holes 
\cite{wald}. In 1974, S.W. Hawking observed that black holes radiate as
a consequence of quantum effects and that the spectrum of this radiation
is thermal \cite{hawking}. Therefore, black holes can evaporate and in order
to understand this process we are necessarily led to a ``marriage"
between quantum mechanics and gravity even in the presence of a
weak gravitational field.  The divergent basic assumptions underlying
these two theories gave birth to a number of serious problems which
hadn't been faced till then. A characteristic paradox which has
not yet been resolved is the ``loss of information". For all such issues,
two-dimensional black holes can be a theoretical lab since they describe
the spherically symmetric sector of the corresponding four-dimensional
geometries. Moreover, in 1992 two-dimensional black hole solutions were
seen as background solutions of effective actions emerging from string
theory \cite{witten,mandal}, leading to a more careful study
of this field  \cite{callan,giddings}.
\par
\noindent
In this work, motivated by the calculation of the Bogoliubov coefficients
in the background of a two-dimensional Schwarzschild
black hole formated by the collapse of conformal matter \cite{giddings},
we calculate the corresponding effect in the background of a primordial
charged blak hole. The paper is organized as follows. In section 1 we
repeat the calculation of Giddings and Nelson in the case of a primordial
black hole. That is we do not invoke the region of the linear dilaton
vacuum and we choose conveniently the ingoing states. The coincidence
with the results of \cite{giddings} proves the correctness of our choice.
In section 2 we extend the analysis of section 1 to the case of a
two-dimensional charged black hole. In particular we calculate the
Bogoliubov coefficients, the resulting thermal distribution and the
corresponding temperature. Our results are, of course, in accordance
with what is known about charged black holes. One of the interesting
outcomes of our
calculation is that it yields regular continuous limit both for the
case of an uncharged and an extremal black hole. Finally we conclude
with a discussion of our results.

\section{``Schwarzschild" Black Hole}
 The Bogoliubov coefficients for the case of
 a ``Schwarzschild" two-dimensional black hole
have been explicitly determined  in \cite{giddings}.
The background of this process is taken to be collapsing matter.
 The formated black hole is
characterized by a thermal distribution of Hawking
radiation at temperature $T=\frac{\lambda}{2\pi}$.
In this section we are going to compute the  Hawking
radiation emitted by a ``Schwarzschild" black hole
without invoking the formation process.
Before proceeding with the calculation of the Bogoliubov
coefficients we will give the line element of the
black hole in various coordinate systems.This is important
since various parts of the calculations that follow
are simplified by the use of different coordinate systems.
\newline
{\bf i) ``Schwarzschild" gauge}
\newline
The two-dimensional dilatonic black hole in the
``Schwarzschild" gauge is characterized by the line element:
\be
ds^2=-g(r)dt^2+g^{-1}(r)dr^2
\label{schdil}
\ee
where the function $g(r)$ is given by:
\be
g(r)=1-\frac{M}{\lambda}e^{-2\lambda r}
\ee
and $0<t<+\infty$, $r_H<r<+\infty$, with $
r_H=\frac{1}{2\lambda}ln(\frac{M}{\lambda})$ the
position of the event horizon of the black hole.
\newline
{\bf ii) Unitary gauge}
\newline
The line element is:
\be
ds^2=-tanh^2(\lambda y)dt^2+dy^2
\label{unidil}
\ee
where the ``unitary" variable $y$ is given by
the following expression:
\be
y=\frac{1}{\lambda}ln[e^{\lambda (r-r_H)} +
\sqrt{e^{2\lambda (r-r_H)}-1}]
\ee
and $0<y<+\infty$.
\newline
{\bf iii) Conformal gauge}
\newline
The line element in this gauge is:
\be
ds^2=(1+e^{-2\lambda x})^{-1}(-dt^2+dx^2)
\label{condil}
\ee
where the variable $x$ is given by:
\be
x=\frac{1}{2\lambda}ln[e^{2\lambda(r-r_H)}-1]
\ee
and $-\infty<x<+\infty$.
\newline
{\bf iv) Asymmetric gauge}
\newline
The corresponding line element is:
\be
ds^2=-\frac{X}{X+1}dt^2+\frac{dX^2}{4\lambda^2X(X+1)}
\label{asdil}
\ee
where the ``asymmetric" variable is given by:
\be
X= e^{2\lambda(r-r_H)}-1
\ee
and $0<X<+\infty$.
\par
\noindent
Considering now a massless scalar field $\Phi$
satisfying the Klein-Gordon equation $\Box \Phi=0$
we have for example in the unitary gauge:
\be
-\frac{1}{tanh^2(\lambda y)} \frac{\partial^2}{\partial t^2}
\Phi (t,y) + \frac{1}{tanh(\lambda y)}
\frac{\partial}{\partial y} [tanh(\lambda y)
\frac{\partial\Phi(t,y)}{\partial y}] =0
\ee
The solution for the modes of the scalar field 
(following E.C. Titchmarsh \cite{titch}) is given by:
\be
\Phi_\omega(t,y)=N(\omega)e^{-i\omega t}
[sinh(\lambda y)]^{i\omega / \lambda},
\label{modesdil}
\ee
where the normalization factor reads:
\be
N(\omega)=\frac{2^{i\frac{\omega}{2\lambda}}}
{\sqrt{4\pi\omega}}
\ee
As it is well known \cite{birrell} the Bogoliubov coefficients
are given by:
\bea
\alpha_{\omega \omega'} &=&
(u_\omega^{out}, u_{\omega'}^{in}) =
-i\int_\Sigma u_\omega^{out}(t,y)\stackrel{\leftrightarrow}{\partial_\mu}
u_{\omega'}^{in*}(t,r)\eta^\mu d\Sigma  
\label{defbog1}\\
\beta_{\omega \omega'} &=&
-(u_\omega^{out}, u_{\omega'}^{in*}) =
i\int_\Sigma u_\omega^{out}(t,y)\stackrel{\leftrightarrow}{\partial_\mu}
u_{\omega'}^{in}(t,r)\eta^\mu d\Sigma
\label{defbog}
\eea
where $ u_\omega^{out}$, ($u_{\omega'}^{in}$) two complete sets of
outgoing (incoming) states and the integration performed on a hypersurface.
For our calculation we take as outgoing states the solutions in
 (\ref{modesdil}) while for incoming states we choose
plane waves

\be
u_\omega^{in}(t,r)=\frac{1}{\sqrt{4\pi\omega}}e^{-i\omega t}
e^{i\omega r}.
\label{plane}
\ee

 Working in the "Schwarzschild" gauge where $d\Sigma=dr$,
 $\eta^{\mu}=(-1,0)$ the future directed timelike vector and identifying
 the variable $r$ in (\ref{plane}) with the one of the corresponding
 gauge we can evaluate the integrals in (\ref{defbog1}, \ref{defbog}).
  A
 lengthy but straightforward calculation \cite{grad} yields for
the Bogoliubov coefficients:
\bea
\alpha_{\omega \omega'} &=&
-\frac{1}{4\pi\lambda}
\sqrt{\frac{\omega'}{\omega-i\epsilon}}2^{i\frac{\omega}{2\lambda}}
\left(\frac{M}{\lambda}\right)^{-i\frac{\omega'}{2\lambda}}
B \left(-i\frac{\omega}{2\lambda}+i\frac{\omega'}{2\lambda}
+\epsilon,1+i\frac{\omega}{2\lambda} \right)
\label{bogoa}\\
\beta_{\omega \omega'} &=&
\frac{1}{4\pi\lambda}
\sqrt{\frac{\omega'}{\omega-i\epsilon}}2^{i\frac{\omega}{2\lambda}}
\left(\frac{M}{\lambda}\right)^{-i\frac{\omega'}{2\lambda}}
B \left(-i\frac{\omega}{2\lambda}-i\frac{\omega'}{2\lambda}
+\epsilon,1+i\frac{\omega}{2\lambda} \right)
\label{bogob}
\eea
where $\epsilon>0$ a positive quantity which is necessary
in order that the expressions in (\ref{bogoa}, \ref{bogob})
be well defined \cite{giddings}, and $B$ stands for the
beta function.
The corresponding probabilities, giving the spectrum of the
black hole radiation are taken by squaring
(\ref{bogoa}, \ref{bogob}) leading to:
\bea
\mid \alpha_{\omega \omega'} \mid ^2 &=& \left(\frac{1}
{4 \pi \lambda} \right)^2 \mid \frac{\omega '}
{\omega - i\epsilon} \mid \mid
B \left(-i\frac{\omega}{2\lambda}+i\frac{\omega'}{2\lambda}
+\epsilon,1+i\frac{\omega}{2\lambda} \right) \mid ^2 \\
\mid \beta_{\omega \omega'} \mid ^2 &=& \left(\frac{1}
{4 \pi \lambda} \right)^2 \mid \frac{\omega '}
{\omega - i\epsilon} \mid \mid
B \left(-i\frac{\omega}{2\lambda}-i\frac{\omega'}{2\lambda}
+\epsilon,1+i\frac{\omega}{2\lambda} \right) \mid ^2
\eea
The above spectrum leads to a thermal distribution
at temperature \cite{birrell} :
\be
T_H=\frac{\lambda}{2 \pi}.
\label{tempa}
\ee

Note that the above result can be reached in any coordinate
system just by taking care of the appropriate Jacobian. The
coincidence with the results of \cite{giddings} points to the
fact  that
the careful use of the plane waves as incoming states is correct
even if we have not the linear dilaton vacuum region which exists
in \cite{giddings}. We can thus proceed with the consideration
of the charged black hole.

\section{Charged Black Hole}

The line element of the charged two-dimensional
black hole is given by (in coordinates corresponding
to the ``Schwarzschild" gauge of the previous section) \cite{kim}:

\be
ds^2 = -g(r)dt^2 + g^{-1}(r)dr^2
\label{linelement2}   
\ee
where
\be
g(r) = 1 - \frac{M}{\lambda} e^{-2\lambda r} +
\frac{Q^2}{4 \lambda ^2} e^{-4\lambda r}
\ee
with $0<t<+\infty$, $r_+<r<+\infty$, $r_+$ being the
future event horizon of the black hole. 
\newline
Following a parametrization analogous to the four-dimensional case the metric function factorizes as:
\be
g(r)=(1-\rho_- e^{-2\lambda r})(1-\rho_+e^{-2\lambda r})
\ee
where
\be
\rho_\pm=\frac{M}{2 \lambda}\pm\frac{1}{2\lambda}
\sqrt{M^2-Q^2}
\label{root2}
\ee
we can recognize immediately the outer event horizon 
$H^+$ placed at the point $r_+=\frac{1}{2\lambda}ln\rho_+$,
while the ``inner" horizon $H^-$ is at the point
$r_-=\frac{1}{2\lambda}ln\rho_-$. In the extremal case ($Q=M$)
the two surfaces coincide in a single event horizon
at the point:
\be
r_H=\frac{1}{2 \lambda}ln \left( \frac{M}{2 \lambda} \right).
\ee
The line element (\ref{linelement2}-\ref{root2}) takes the 
following forms in the different coordinate systems previously considered:
\newline
{\bf i) Unitary gauge}
\newline
The line element reads
\be
ds^2=-\frac{\mu^2 sinh^2(\lambda y) cosh^2 (\lambda y)}
{[\mu sinh^2(\lambda y) +1]^2}dt^2 + dy^2
\ee
where the unitary variable is given by
\be
y=\frac{1}{\lambda}ln\left[
\sqrt{\frac{1}{\mu}(e^{2\lambda(r-r_+)}-1)} +
\sqrt{\frac{1}{\mu}(e^{2\lambda(r-r_+)}-1)+1} \right]
\ee
and $0<y<+\infty$, while $\mu = 1-\frac{\rho_-}{\rho_+}$.
\newline
{\bf ii) Asymmetric gauge}
\newline
In this coordinate system the line element reads:
\be
ds^2=-\frac{X(X+\mu)}{(X+1)^2} dt^2 +
\frac{dX^2}{4 \lambda^2X(X+\mu)}
\ee
where $X=e^{2\lambda(r-r_+)}-1=\frac{e^{2\lambda r}}{\rho_+}-1$,
and $0<X<+\infty$.
\newline
{\bf iii) Conformal Gauge}
\newline
The line element of the two-dimensional spacetime is:
\be
ds^2 = \frac{X(x)[X(x)+\mu]}{[X(x)+1]^2}(-dt^2+dx^2)
\ee
where the ``conformal" variable $x$ is given by:
\be
x=\frac{1}{2\lambda} ln \left[X^{\frac{1}{\mu}}(
X+\mu)^{1-\frac{1}{\mu}} \right]
\ee
and $-\infty<x<+\infty$.
\newline
Considering again a massless scalar field $\Phi$ satisfying the
Klein-Gordon equation $\Box \Phi=0$ in the conformal gauge, which
in this case is more convenient  for simplicity reasons,
 we get
the solution for the modes of the scalar field :
\be
\Phi_\omega(t,x)=N(\omega)e^{-i\omega t}e^{i\omega x}
\ee
where the normalization factor reads :
\be
N(\omega)=\frac{1}{\sqrt{(2\pi)(2\omega)}}
\ee
\par
\noindent
Now we can follow the analysis of previous section. Our incoming
states are again plane waves, with the space
variable identified to the variable of the corresponding
"Schwarzchild" gauge. In this case
it turns out that the asymmetric gauge is more convenient
for evaluating the integrals in  (\ref{defbog1}), (\ref{defbog});
one needs only to use the
appropriate Jacobian. Skipping the details of the calculation,
 the Bogoliubov coefficients for the case of
two-dimensional charged black hole are found to be :

\bea
\alpha_{\omega \omega'} = -\frac{1}{4\pi\lambda}
\sqrt{\frac{\omega'}{\omega-i\epsilon}}\rho_+^{
-i\frac{\omega'}{2\lambda}} 
B\left[
-i\frac{\omega}{2\lambda}+i\frac{\omega'}{2\lambda}+
\epsilon ,1+i\frac{\omega}{2\lambda\mu} \right]\times  \nn \\
_2F_1 \left[ -i\frac{\omega}{2\lambda} +
i\frac{\omega}{2\lambda\mu} ,
-i\frac{\omega}{2\lambda}+ i\frac{\omega'}{2\lambda};
-i\frac{\omega}{2\lambda}+i\frac{\omega'}{2\lambda}+
i \frac{\omega}{2\lambda\mu}+1 ;1-\mu \right]  
\label{boga}\\
\beta_{\omega\omega'} = \frac{1}{4\pi\lambda}
\sqrt{\frac{\omega'}{\omega-i\epsilon}}\rho_+^{
i\frac{\omega'}{2\lambda}}
B\left[
-i\frac{\omega}{2\lambda}-i\frac{\omega'}{2\lambda}+
\epsilon , 1+i\frac{\omega}{2\lambda\mu} \right] \times \nn \\
_2F_1 \left[ -i\frac{\omega}{2\lambda} +
i\frac{\omega}{2\lambda\mu} ,
-i\frac{\omega}{2\lambda}- i\frac{\omega'}{2\lambda} ;
-i\frac{\omega}{2\lambda}-i\frac{\omega'}{2\lambda}+
i \frac{\omega}{2\lambda\mu}+1 ; 1-\mu \right]
\label{bogb}
\eea
\par
\noindent
Squaring we get the probabilities for particle emission:
\bea
|\alpha_{\omega \omega'}|^2 = \left( \frac{1}{4\pi\lambda} \right)^2
|\frac{\omega'}{\omega-i\epsilon}| 
| B\left[
-i\frac{\omega}{2\lambda}+i\frac{\omega'}{2\lambda}+
\epsilon ,1+i\frac{\omega}{2\lambda\mu} \right]|^2 \times  \nn \\
|_2F_1 \left[ -i\frac{\omega}{2\lambda} +
i\frac{\omega}{2\lambda\mu} ,
-i\frac{\omega}{2\lambda}+ i\frac{\omega'}{2\lambda};
-i\frac{\omega}{2\lambda}+i\frac{\omega'}{2\lambda}+
i \frac{\omega}{2\lambda\mu}+1 ;1-\mu \right]|^2 
\label{bogfa} \\
|\beta_{\omega\omega'}|^2 = \left( \frac{1}{4\pi\lambda} \right)^2
|\frac{\omega'}{\omega-i\epsilon}|
| B\left[
-i\frac{\omega}{2\lambda}-i\frac{\omega'}{2\lambda}+
\epsilon , 1+i\frac{\omega}{2\lambda\mu} \right]|^2 \times \nn \\
| _2F_1 \left[ -i\frac{\omega}{2\lambda} +
i\frac{\omega}{2\lambda\mu} ,
-i\frac{\omega}{2\lambda}- i\frac{\omega'}{2\lambda} ;
-i\frac{\omega}{2\lambda}-i\frac{\omega'}{2\lambda}+
i \frac{\omega}{2\lambda\mu}+1 ; 1-\mu \right]|^2
\label{bog} 
\eea
which are useful for the evaluation of the Hawking radiation for a 
two-dimensional charged black hole.
\par
It is noteworthy to observe that
if we set $\mu =1$, (in other words $\rho_- =0$, i.e. the black hole
 has no charge ($Q=0$), then the expressions (\ref{boga}) and (\ref{bogb})
  coincide with expressions (\ref{bogoa}) and (\ref{bogob}) respectively.
   It is also very interesting to observe the consequences
of the presence of the electric charge to the ``equilibrium" state in
 which the black hole has settled down and more specifically the relation
 between the electric charge and the temperature. If we wish to count the
 particles which are detected in the case of ``Schwarschild" black hole and
  charged black hole, we have to evaluate the following quantity :
\be
_{in}\langle{0}\vert{N_\omega^{out}}\vert{0}\rangle_{in}=
\int_{0}^{+\infty}d\omega'{\vert{\beta}_{\omega\omega'}\vert}^{2}
\label{integral}
\ee
\par
\noindent
For our convenience and without loss of generality,
we set $\lambda=\frac{1}{2}$
and the equation (\ref{bog}) becomes :
\bea
|\beta_{\omega\omega'}|^2 = \left( \frac{1}{2\pi} \right)^2
|\frac{\omega'}{\omega-i\epsilon}|
| B\left[
-i\omega-i\omega'+
\epsilon , 1+i\frac{\omega}{\mu} \right]|^2 \times \nn \\
| _2F_1 \left[ -i\omega+
i\frac{\omega}{\mu} ,
-i\omega- i\omega';
-i\omega-i\omega'+
i \frac{\omega}{\mu}+1 ; 1-\mu \right]|^2 
\eea
Plotting the quantity (\ref{integral}) as a function of
frequency $\omega$ of the outgoing modes for different values
 of the parameter $\mu$ (which denotes the presence of the electric
  charge and the ratio between mass $M$ and charge $Q$), we get figure 1.

\begin{figure}[h]
\begin{center}
\includegraphics[width=10cm,height=7cm]{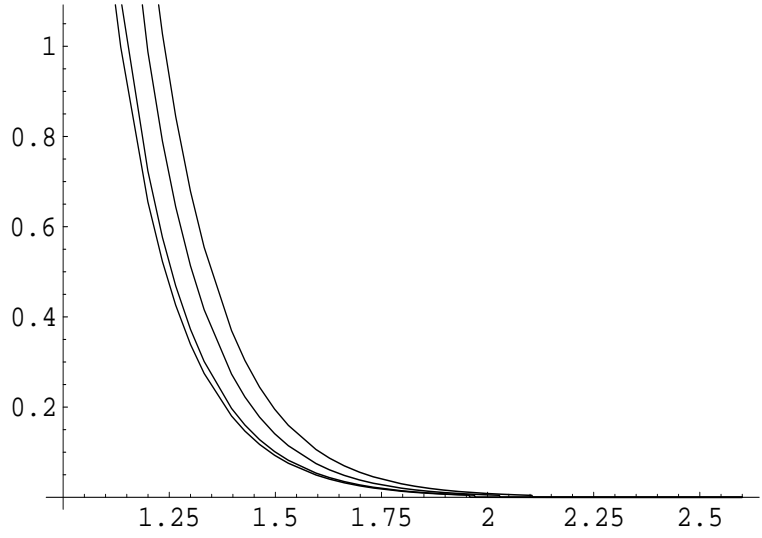}
\caption{Diagrams obtained from the integration
 of the quantity (\ref{integral}) over the variable
 $\omega'$ for $\mu=1$, $\mu=0.9$, $\mu=0.7$ and $\mu=0.5$ respectively.}
\end{center}
\end{figure} 
\noindent
It is obvious from the above graph that as $\mu$ decreases, i.e.
the ratio $\frac{Q}{M}$ increases, the graph ``falls down". This
means that the corresponding black holes have lower
temperature.
This result can also be reached considering the definition
 \cite{birrell}:
\be
T_H=\frac{\kappa}{2\pi}\hspace{.2in}and\hspace{.2in}
{\kappa}=\frac{1}{2}\left. 
\frac{\partial g(r)}{\partial r}\right|_{r=r_H}
\ee
which leads to the following expression for the Hawking temperature :
\be
T_H=\frac{\lambda}{2\pi}{\mu}
\label{tempb}
\ee
\par
\noindent
Equation (\ref{tempb}) is the ``Reissner-Nordstrom" analog of 
equation (\ref{tempa}) holding for the ``Schwa\-rzschild" case.
When the parameter $\mu$ approaches 1, i.e. the electric charge is
zero ($Q=0$),
we get that (\ref{tempb}) goes over to (\ref{tempa}).When the parameter
$\mu$ approaches zero, i.e. the electric charge is equal to mass
($Q=M$, extremal case), then the temperature is zero ($T=0$),
i.e. the black hole ``freezes" completely.
\noindent
This known result
\cite{frolov,lee,anderson} appears here as the regular
($\mu \rightarrow 0$ or $\mu \rightarrow 1$)
limit of the more general case. This can be seen either from figure 1,
or analytically working the limiting cases of
(\ref{bogfa}), (\ref{bog}). The zero charge case has already been
discussed. Furthermore the coefficients in  the abovementioned
expressions
go to those of the flat space
 as $\mu \rightarrow 0$
 indicating the ``freezing"
 of the extremal black hole. Namely
$|\alpha_{\omega \omega'}|^2 \rightarrow \delta_{\omega \omega'}$ and
$|\beta_{\omega \omega'}|^2 \rightarrow 0$, for positive to positive
frequency transitions.

\section{Discussion}
In this work, we have explicitly calculated the Bogoliubov coefficients
for a background which is an already formated (primordial) two-dimensional
 black hole with : (a) mass $M$  (``Schwarzschild" case), (b) mass $M$ and
 electric charge $Q$ (``Reissner-Nordstrom" case). We have evaluated
 graphically their corresponding thermal spectrums and we have found
 an explicit expression for their temperatures when the system has
  settled down to an ``equilibrium" state. We have concluded that the
  temperature of the Hawking radiation for the two-dimensional charged
   black hole is a function of charge and
mass as it is in the case of four-dimensional black hole, in
contradistinction to the two-dimensional ``Schwarzschild"
case where temperature does not depend on mass. We have shown that
as the mass of the two-dimensional charged black hole decreases,
 the temperature of Hawking radiation also decreases till the moment
 when the mass becomes equal to the charge (extremal case) at which
 moment the two-dimensional black hole ``freezes" completely. This
 denotes that the extremal black holes can be quantum mechanically
 a stable ending point for the black holes during the process of their
 evaporation.
 \par
 \noindent
 It is worthexploring how the analytical results discussed in
 this work can be applied for the study of thermodynamic
 properties of the two-dimensional black holes. It is also
 interesting to see how the calculations performed for the
 two-dimensional case can be extended for the higher
 dimensional geometries, and/or include the other charge
 a black hole can carry (angular momentum).
 We hope to return to this issues in a future work.

 \end{document}